\documentclass[pre,aps,showpacs,amsmath,amssymb,twocolumn]{revtex4-1}

\usepackage{graphicx}

\begin{document}

\preprint{PREPRINT}

\title{Replica Exchange Monte Carlo applied to Hard Spheres}

\author{Gerardo Odriozola} 

\affiliation{Programa de Ingenier\'{\i}a Molecular, Instituto
Mexicano del Petr\'{o}leo, L\'{a}zaro C\'{a}rdenas 152, 07730
M\'{e}xico, D. F., M\'{e}xico}

\date{\today}
\begin{abstract}
In this work a replica exchange Monte Carlo scheme which considers
an extended isobaric-isothermal ensemble with respect to pressure
is applied to study hard spheres (HS). The idea behind the
proposal is expanding volume instead of increasing temperature to
let crowded systems characterized by dominant repulsive
interactions to unblock, and so, to produce sampling from disjoint
configurations. The method produces, in a single parallel run, the
complete HS equation of state. Thus, the first order fluid-solid
transition is captured. The obtained results well agree with
previous calculations. This approach seems particularly useful to
treat purely entropy-driven systems such as hard body and
non-additive hard mixtures, where temperature plays a trivial
role.
\end{abstract}


\maketitle

\section{Introduction \label{introduction}}

The replica exchange Monte Carlo (REMC) method
\cite{Lyubartsev91,Marinari92}, also called parallel tempering
\cite{Yan99}, was derived to achieve good sampling of systems that
present a free energy landscape with many local minima
\cite{Bittner08,Frenkel}. It consists on simulating several
replicas of the same system at different thermodynamic states, and
allowing for replica exchanges (swap moves). Thus, it is possible
to implement an ergodic walk through free energy barriers
connecting disjoint configuration subspaces by defining a set of
close enough thermodynamic states. Although it has been developed
at the end of the last century \cite{Lyubartsev91,Marinari92}, its
acceptance is already high due to its clearness, simplicity, and
its wide applicability. Proof of that is its employment to find
zeolite structures \cite{Falcioni99}, to study different
conformations of proteins\cite{Hernandez08}, and to access phase
equilibrium of many single and multicomponent systems
\cite{Fiore08,Imperio06,Arnold08}.

Most frequently, the REMC technique is employed to sample an
extended canonical ensemble in temperature. Thus, those replicas
having larger temperatures are capable of escaping from local free
energy minima, where the pair potential attraction of the
constituting particles plays a key role. When the free energy
minima are mainly dictated by the entropic term, i. e., by the
excluded volume repulsive interactions
\cite{Fortini06,Odriozola08,Jimenez08}, enlarging the temperature
has a small effect. In other words, the benefits of the method
become restricted. This is especially true when dealing with hard
body systems (purely entropy-driven systems) such as hard spheres
(HS), rods, plates, polymers, and non-additive hard mixtures,
since they constitute limiting cases where the pair interactions
are repulsive only and the temperature plays a trivial (null)
role. Thus, it is not very surprising that the REMC technique has
not been applied yet to this kind of systems. To do this, an
alternative would be performing the ensemble extension in pressure
instead of temperature, to provide the particles more freedom to
rearrange as the volume expands. This idea is tested in this work
for HS.

It is well known that fluid-solid transitions represent a
challenge for computational science \cite{Wilding00,Noya08}. Most
techniques which properly work for accessing liquid-gas
transitions have problems at very high densities\cite{Frenkel}.
Therefore, the freezing and melting points are at least difficult
to determine \cite{Wilding00}. Indeed, for the HS model,
simulations have recently produced an accurate determination of
the freezing and melting point theoretically reported in the
sixties \cite{Wilding00,Noya08}. That is despite the intense study
of HS through the past decades, and the fact that the HS model was
one of the first systems ever studied by computer simulations
\cite{Rosenbluth54,Wood57,Alder57}. Additionally, the HS model
shows a high density metastable branch ending at the random close
package density \cite{Rintoul96}, which adds difficulty for
sampling from equilibrium.

The aim of this study is to show that the REMC can be successfully
applied to study hard body systems. Hence, the REMC is used by
performing a NPT ensemble extension on pressure and applied to HS.
The paper is structured as follows. Sec.~\ref{introduction} is
this brief introduction. Sec.~\ref{method} describes the employed
algorithm. Results are given in Sec.~\ref{results}. Finally, in
Sec.~\ref{conclusions} conclusions are drawn.

\section{Numerical Method}
\label{method}

As mentioned, in the parallel tempering scheme $n_r$ identical
replicas are considered, each following a typical canonical
simulation. However, a different temperature is set for each one
of them. Thus, an extended ensemble can be defined so that its
partition function is $Q_{extended}=\prod_{i=1}^{n_r}Q_{NVTi}$,
being $Q_{NVTi}$ the partition function of ensemble $i$ at
temperature $T_i$, number of particles $N$, and volume $V$. The
existence of this extended ensemble justifies the introduction of
swap trial moves between any two ensembles (each ensemble is
sampled by only one replica at a time), whenever the detail
balance condition is satisfied. If all
$(i,T_i)(j,T_j)\!\!\rightarrow\!(j,T_i)(i,T_j)$ swap trials have
the same a priori probability of being performed, the swap
acceptance probability becomes
\begin{equation}\label{accT}
P_{acc}\!=\! min(1,\exp[( \beta_j - \beta_i)(U_{i}-U_{j})])
\end{equation}
where $\beta_i=1/(k_BT_i)$ is the reciprocal temperature of
replica $i$, $k_B$ is the Boltzmann's constant, and $U_i$ is the
energy of replica $i$. Hence, by introducing these swap trials, a
particular replica seals through many temperatures allowing it to
overcome free-energy barriers. Additionally, sampling on
particular ensembles is not disturbed but enriched by the
different contributions of the $n_r$ replicas.

For studying systems where excluded volume interactions dominate,
it may be convenient to allow the replicas to expand for
destroying any local order. Additionally, in the case of a HS
system (or any other purely entropy-driven model) an extended
ensemble in temperature is pointless, since this variable does not
affect the system structure. For that purpose, the extended
ensemble is defined as $Q_{extended}=\prod_{i=1}^{n_r}Q_{NTPi}$,
being $Q_{NTPi}$ the partition function of the isobaric-isothermal
ensemble of system $i$, at pressure $P_i$, fixed temperature $T$,
and number of particles $N$ (note that the extension is in
pressure; an isobaric-isotermal extension in temperature applied
to a Lennard Jones system is given by Okabe et. al.
\cite{Okabe01}). This extended ensemble can be sampled by
performing standard $NTP$ simulations on each replica, which
implies typical particle displacement trials and volume change
trials. Notwithstanding, the sampling can be significantly
improved by introducing swap trials between neighboring ensembles.
Again, the only restriction is that the detail balance condition
must prevail to guaranty the correct sampling. One way of
achieving this is by setting equal all a priori probabilities of
choosing the different adjacent pairs of replicas, and accounting
for the following acceptance probability
\begin{equation}\label{accP}
P_{acc}\!=\!min(1,\exp[\beta(P_i- P_j)(V_i-V_j)])
\end{equation}
where $V_i-V_j$ is the volume difference between replicas $i$ and
$j$. It should be noted that the ensemble extension in pressure
leads to a simple acceptance rule where energy terms vanish.

For $(P_i-P_j)(V_i-V_j)\geq 0$, $P_{acc}=1$ and so, the acceptance
rule tends to order the replicas by volume size (lower volumes at
higher pressures). For $(P_i-P_j)(V_i-V_j) < 0$, $P_{acc}$ depends
on the absolute value of the pressure differences of the adjacent
ensembles, $\beta|P_i- P_j|$. A decrease of $\beta|P_i- P_j|$
leads to a larger acceptance probability. Consequently, adjacent
pressures should be close enough to provide large exchange
acceptance rates between neighboring ensembles. This is
particularly important where a phase transition takes place
(characterized by large $|V_i-V_j|$), which generally leads to a
bottleneck of the swap acceptance rate. Additionally, the swap
acceptance rate also depends on the system size. Larger system
sizes produce narrower distribution of densities (volumes) for a
given pressure, providing smaller overlap regions between adjacent
ensembles. Hence, a larger system size leads to a decrease of the
swap acceptance probability. Finally, in order to take a good
advantage of the method, the replica at the lowest pressure must
assure large jumps in configuration space, so that the higher
pressure ensembles can be sampled from disjoin configurations.

In this work, $n_r=70$ cubic boxes of initial side $L$ are
considered. These boxes are filled by randomly placing $N$ hard
spheres of diameter $\sigma$. The initial density, $\rho=N \pi
\sigma^3/(6L)$, is set to 0.30 for all replicas. A geometrically
increasing pressure, $\beta P$, is set from approximately 2 to 100
$\sigma^{-3}$, and arbitrarily assigned to the replicas. Where the
fluid-solid transition is expected, intermediate pressures are
added (the total number of replicas equals the number of different
pressures). An optimal allocation of replicas should lead to a
constant swap acceptance probability for all pair of adjacent
ensembles \cite{Rathore05}. Two experiments were done, one with
$N=32$ and the other with $N=108$. The simulation starts by
following the trial moves above described (see the appendix for
details).

Sampling consists on measuring densities, radial distribution
functions, average number of neighbors, and the order parameter
$Q_6$, as a function of the pressure. The average number of
neighbors, $N_n$, is computed accounting for all pairs having a
center-center distance smaller than $1.2\sigma$ (the vectors
joining the centers of these pairs are named bonds). The order
parameter $Q_6$ is defined as \cite{Steinhardt83,Rintoul96}
\begin{equation}\label{Q6}
Q_6=\left(\frac{4\pi}{13}\sum_{m=-6}^{m=6}|<\!Y_{6m}(\theta,
\phi)\!>|^2\right)^{1/2}
\end{equation}
where $<\!Y_{6m}(\theta, \phi)\!>$ is the average over all bonds
and configurations of the spherical harmonics of the orientation
angles $\theta$ and $\phi$ (these are the polar angles of the
bonds measured with respect to any fixed coordinate system, since
$Q_6$ is invariant). $Q_6$ should go to zero for a completely
random system of a large number of points, following
$1/\sqrt{NN_n/2}\pm 1/\sqrt{13NN_n}$ \cite{Rintoul96}.

\section{Results}
\label{results}

\begin{figure}
\resizebox{0.48\textwidth}{!}{\includegraphics{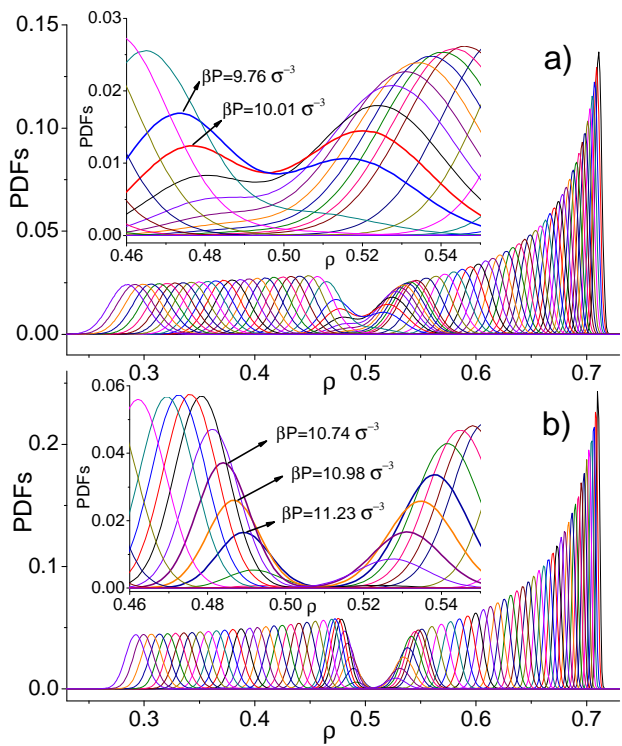}}
\caption{\label{PDFs} Probability density functions, PDFs, to find
a replica at a given density, $\rho$. The 70 different curves
correspond to the different assigned pressures. Fig. a)
corresponds to $N=$32 and Fig. b) to $N=$108. Both insets zoom in
the corresponding data. }
\end{figure}

Figure \ref{PDFs} a) shows the probability density functions,
PDFs, to find a replica at a given density for all pressures and
for $N=32$. The 70 PDFs correspond to the different assigned
pressures. In general, the PDFs are bell-shaped and centered on a
maximum which location depends on the assigned pressure. The
leftmost curve corresponds to the lowest pressure (2.16
$\beta^{-1} \sigma^{-3}$) and the rightmost to the highest one
(100 $\beta^{-1} \sigma^{-3}$). As pressure increases, the curves
narrow and shift to the right producing larger densities (the
narrowing is very pronounced for high pressures). The exception
occurs for densities close to $0.5$, where the PDFs split yielding
bimodal distributions. At $\rho \simeq 0.5$, the replicas produce
few configurations, and the bimodals yield a local peak below 0.49
and another above 0.51. Thus, a jump on density from $\rho_f
=0.474$ to $\rho_s=0.520$ is produced for $\beta P = (9.95 \pm
0.10) \sigma^{-3}$, pointing out the well known HS fluid-solid
transition. The inset of figure \ref{PDFs} a) zooms in the density
region around 0.5, where the PDFs are much clearly seen. There it
is shown the pressures that correspond to the PDFs which are
closer to the transition.

The PDFs obtained for $N=108$ are shown in figure \ref{PDFs} b).
As expected, similar trends are seen. That is, PDFs are
bell-shaped, they narrow and shift toward larger densities for
increasing pressure, and they turn bimodal for densities close to
0.5. Nevertheless, PDFs are higher (approximately two times
higher) and (consequently) narrower than for $N=32$. Also the
bimodal distributions become sharper producing interpeak regions
rarely visited by the replicas. In fact, for $\rho \simeq 0.508$
the PDFs are practically zero. In other words, the HS fluid-solid
transition turns more evident by increasing the system size. As in
figure \ref{PDFs} a), the inset of figure \ref{PDFs} b) zooms in
the corresponding PDFs. From there it can be estimated the
transition occurring at $\beta P = (10.99 \pm 0.10) \sigma^{-3}$
with $\rho_f=0.487$ and $\rho_s=0.538$. Thus, the transition
occurs at a higher pressure and shifts to larger densities for
increasing the system size. The gap between the fluid and solid
densities also enlarges.

\begin{figure}
\resizebox{0.40\textwidth}{!}{\includegraphics{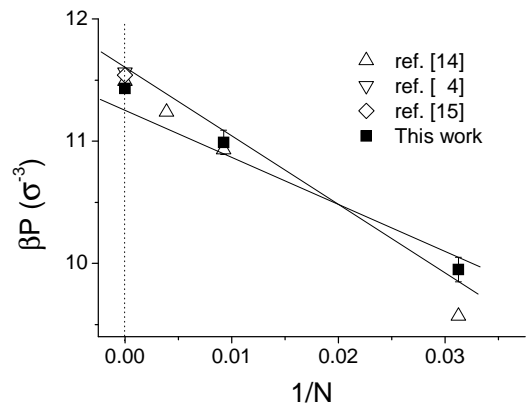}}
\caption{\label{comparison} $\beta P$ as a function of the inverse
of the system size, $1/N$. Solid symbols correspond to this work
results. The solid symbol at $1/N=0$ is an extrapolation of the
data and the solid lines are drawn to estimate the corresponding
error. Open symbols are values reported by different authors.}
\end{figure}

The data obtained for small $N$ values can be extrapolated to
estimate the HS bulk coexistence pressure, fluid density, and
solid density. These are $\beta P_{tr} = (11.43 \pm 0.17)
\sigma^{-3}$, $\rho_f=$0.492 $\pm$ 0.004, and $\rho_s=$0.545 $\pm$
0.004, respectively. These values are in good agreement with
previous calculations \cite{Frenkel,Wilding00,Noya08}. Figure
\ref{comparison} shows the extrapolation for the coexistence
pressure, and a comparison with data reported by different
authors. As can be seen, the obtained agreement is good,
suggesting that the RECM method works properly for capturing the
HS fluid-solid transition.

\begin{figure}
\resizebox{0.48\textwidth}{!}{\includegraphics{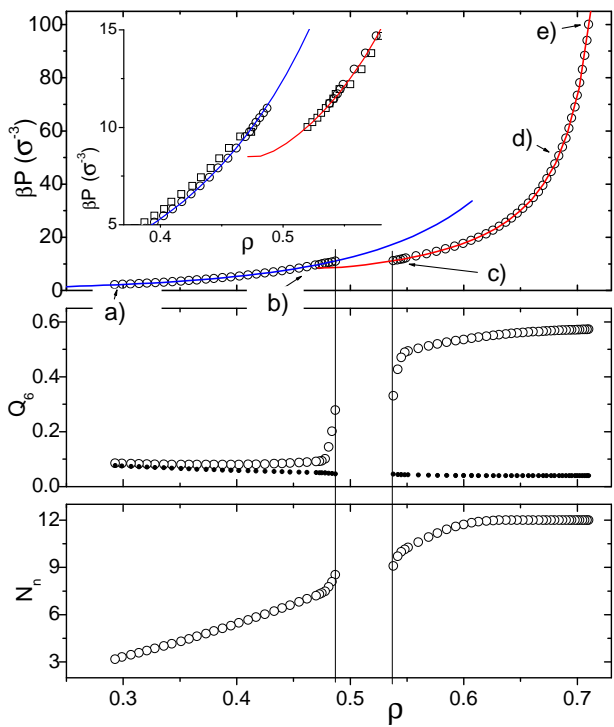}}
\caption{\label{transition} Topmost; HS equation of state
(pressure as a function of the corresponding most frequent
density) for $N=$108 ({\tiny $\bigcirc$} symbols). The red line
corresponds to the HS fluid equation of state of Speedy
\cite{Speedy97} and the blue one to the HS face cubic centered
equation of state of the same author \cite{Speedy98}. Inset; zoom
in of the same plot, where {\tiny $\square$} symbols were added
corresponding to $N=$32. Middle plot; Order parameter, $Q_6$, as a
function of $\rho$ ({\tiny $\bigcirc$} symbols), and the
corresponding value for a completely random system of points
$1/\sqrt{NN_n/2}$ (small bullets). Bottommost; Number of first
neighbors, $N_n$, as a function of $\rho$. }
\end{figure}

The topmost plot of figure \ref{transition} is built by plotting
the pressure as a function of the most frequent density for
$N=108$. It is also shown as a red line a Pad\'{e} approximation
to data obtained from the HS fluid state \cite{Speedy97}, and as a
blue line a fit to the HS face cubic centered (FCC) solid state
\cite{Speedy98}, both data series obtained by means of
simulations. As an inset, it is shown a zoom in of the same data
for the coexistence, where there were added the data obtained for
$N=32$. Both curves here reported, for the fluid and solid states,
well agree with the equation of state given by Speedy. This
confirms the good behavior of the REMC ensemble extension on
pressure. Nevertheless, there is a slight deviation from the FCC
curve of Speedy close to the transition. This may signal the
presence of hexagonal close-packed (HCP) arrangements and even
hybrid FCC-HCP structures.

The middle and bottommost plots of figure \ref{transition} show
the order parameter, $Q_6$, and the number of first neighbors,
$N_n$, as a function of $\rho$, respectively. The middle plot also
shows as bullets the value of $Q_6$ for completely
space-uncorrelated particles. As expected, $Q_6$ is small for the
fluid region, pointing out the practical absence of angular order.
However, it is always somewhat larger than the value of $Q_6$ for
a random system. The difference between these two values
diminishes for decreasing $\rho$. On the other hand, $Q_6$ reaches
$0.5732$ for $\beta P = 100 \sigma^{-3}$, which is slightly lower
than the $Q_6$ value of the FCC arrangement, $0.5745$, and well
above the corresponding value of the HCP structure, $Q_6=0.4848$.
This signals that only replicas approaching the FCC lattice are
allowed for the highest applied pressures. This is not surprising
since 108 identical spherical particles can be perfectly packed on
a cubic box on a FCC lattice, but cannot on a HCP lattice. Thus,
the system is being forced to promote FCC over HCP at high
pressures. For lower but still over the coexistence pressures,
$Q_6$ is close to 0.5, suggesting that both lattices and their
hybrids contribute to the average. It should be noted that $Q_6$
sharply increases at the fluid-solid transition. Thus, it can be
employed to detect any trace of local angular order. This was
shown to be much more reliable than the radial distribution
function peak that develops close to 1.5$\sigma$ \cite{Rintoul96}.
Finally, $N_n$ monotonically increases with $\rho$. It also shows
a sharp increase at the coexistence, although less pronounced than
for $Q_6$. At large densities $N_n$ reaches 12, which is the
largest possible HS coordination number, as it is well known.

\begin{figure}
\resizebox{0.5\textwidth}{!}{\includegraphics{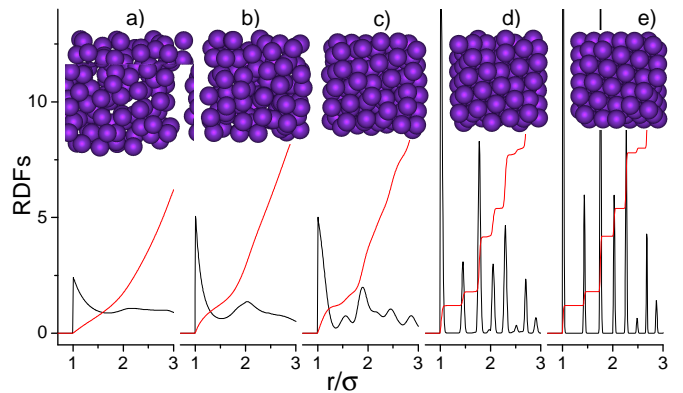}}
\caption{\label{gdrs} Radial distribution functions (black lines)
and their integrals (red lines) for cases a), b), c), d), and e),
as shown in figure \ref{transition}, from left to right. The
insets are the corresponding snapshots. Integrals (red lines) are
scaled by a factor $1/10$.  }
\end{figure}

The radial distribution functions (RDFs) and their corresponding
integrals for cases a), b), c), d), and e) pointed out in figure
\ref{transition} are plotted in figure \ref{gdrs}. Cases a) and b)
correspond to the fluid phase, and the other three correspond to
the solid phase. Case a) shows the typical low density liquid
structure, where a relatively small contact value is developed and
the second shell of neighbors is poorly seen. As the density
increases, case b), the RDF shows a larger contact value (two
times larger than case a)), and a well defined second shell of
neighbors. This case corresponds to a liquid close to the
coexistence. Slightly above the coexistence, the RDF looks like
case c). Here a small peak appears at $r/\sigma \simeq 1.5$,
whereas the valley in-between this peak and the contact one
deepens. Other peaks also form at larger distances. For density
values close to $0.68$, case d), the RDF develops the full
character of a crystal. That is, peaks are very high and narrow,
and valleys turn practically zero. The integral (red line) of this
case highlights this fact, since it shows a step-like behavior.
The first step reaches 12, pointing out the first shell
coordination number (integral of the peak at $r/\sigma=1$), the
second yields 18 ($r/\sigma=\sqrt{2}$), the third 42
($r/\sigma=\sqrt{3}$), and the fourth 54 ($r/\sigma=2$), what
corresponds to the FCC structure. Nonetheless, a small shoulder
appears at the left of the fourth peak ($r/\sigma=\sqrt{11/3}
\simeq 1.91$), suggesting the existence of few configurations
having a HCP structure. For larger densities, case e), this
shoulder disappears and a practically pure FCC RDF is observed.

\section{Conclusions}
\label{conclusions}

This work shows that a replica exchange Monte Carlo scheme can be
successfully applied to study hard spheres at high densities. For
that purpose, an extension of the isobaric-isothermal ensemble
with respect to pressure is used. The algorithm employs standard
particle trial displacements and volume changes together with
replica exchanges (swap moves). These easy to implement trials are
shown to be enough for capturing the fluid-solid transition of
hard spheres and the solid equilibrium branch for small systems.
The obtained results well agree with previous calculations. The
principal idea behind this scheme is to increase the particles
mobilities by decreasing the pressure (expanding the volume), so
that systems characterized by large excluded volume contributions
are able to visit disjoint configurations of configurational
space. This approach seems particularly useful to deal with purely
entropy-driven systems such as hard body and non-additive hard
mixtures, where temperature plays a trivial role.

\section{Appendix- Simulation Details} \label{appendix}

Once the $n_r=70$ boxes are filled with the $N$ spheres and are
assigned the corresponding different pressures, the algorithm
starts performing the trials. As mentioned, they are: particle
displacements, volume changes, and swap moves. The probability for
selecting a particle displacement trial (in any of the $n_r$
boxes), $P_d$, is fixed to $P_d=n_rN/(n_r(N+1)+w(n_r-1))$. The
probabilities for selecting a volume change trial, $P_v$, and a
swap trial, $P_s$, are $P_v=n_r/(n_r(N+1)+w(n_r-1))$ and
$P_s=w(n_r-1)/(n_r(N+1)+w(n_r-1))$. Here, $w$ is a weight factor
fixed to $1/50$. Additionally, the probability of performing a
particle displacement trial and a volume change trial in a replica
enlarges as it is closer to the fluid-solid transition pressure.
These probabilities are 10 times larger for the central replica
than for those having the highest and the lowest pressures. All
particles of a given replica have the same a priori probability of
being selected to perform a displacement trial. The same is true
for selecting a pair of adjacent replicas to attempt a swap move
(there are $n_r-1$ pairs). Thus, a random number homogeneously
distributed in [0,1] is generated in order to determine the type
of trial to be performed. In case of selecting a particle
displacement, the algorithm provides the replica and the particle
for applying the trial. In case of a volume change trial, it
identifies the replica; and in case of a swap trial, the algorithm
gives the adjacent replicas to apply it. Next, another random
number is generated to produce a second trial. If these two trials
are independent the one another (for instance, they are particle
trials on different replicas) the algorithm generates a third
trial (note that these trials are not being applied yet). This
procedure is repeated until the last trial cannot be performed
independently of the others (for instance, a particle displacement
trial on a replica in which a volume change trial must be
previously performed). This way, the algorithm have randomly
selected a given number of independent trials to be applied on the
replicas. Immediately after, the algorithm parallelizes (in two
threads, since a dual core desktop is used), and the trials are
done. The last generated trial (which was not yet performed)
becomes now the first trial to be applied on the following series
of trials. This procedure is followed to strictly preserve the
detail balance condition (to build a symmetric transition matrix)
while performing a parallelization. Verlet lists are employed for
saving CPU time (note that the saving can be quite large since
replicas at high pressures rarely update their lists).

\begin{figure}
\resizebox{0.48\textwidth}{!}{\includegraphics{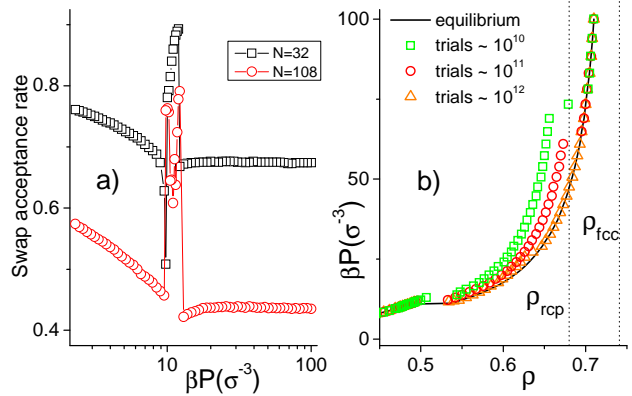}}
\caption{\label{ar-ev} a) Swap acceptance rate as a function of
the ensemble pressure. b) Evolution of the densities with the
number of trials during the initializing procedure ($N=108$). As a
reference, the data obtained from equilibrium are added as a solid
line. The dotted lines show the random close packing density as
obtained from the square symbols and the face cubic centered
density.  }
\end{figure}

Sampling is not performed for the first $3.0 \times 10^{12}$
trials (initializing procedure). During this process, the maximum
displacements of particles and maximum volume changes for each
pressure are tuned to yield acceptance rates close to 0.4. Thus,
particle maximum displacements and maximum volume changes of
ensembles having high pressures turn smaller than those associated
to ensembles having low pressures. Once this is done, all maximum
displacements (one for each pressure) and maximum volume changes
(also one for each pressure) become fixed. $1.0 \times 10^{13}$
trials are then employed to yield the data shown in the body of
the article (equilibrium sampling).

The acceptance rates obtained for the swap trials are shown in
Fig.~\ref{ar-ev} a) as a function of the pressure, $\beta P$. As
expected, the acceptance rates for the smaller system ($N=32$)
are, in general, larger than the ones obtained for the larger
system ($N=108$). This is a consequence of the larger overlaps of
the distributions. In both cases the values are above the
recommended acceptance rate of 0.2 \cite{Rathore05}. For $\beta P
> \beta P_{tr} $, the acceptance rate is practically constant. On
the contrary, for $\beta P < \beta P_{tr}$, the acceptance rate
increases with decreasing $\beta P$. This means that for the fluid
region, $\beta P$ should be reduced more than geometrically for
optimization purposes. For $10 \lesssim \beta P \lesssim 12$, the
acceptance rate increases. This is due to the fact that smaller
pressure differences are set between the adjacent ensembles to
compensate the natural decrease of the acceptance rate at the
fluid-solid transition. Note that more than the necessary replicas
are added in order to decrease the error of the coexistence
pressure (the natural decrease of the acceptance rate is
overcompensated). In addition, this study is focussed on yielding
a detailed sampling of a large $\beta P$ range. To acquire
equilibrium data from a high pressure system only, many fewer
replicas would be required (the optimal swap acceptance rate is
close to $20\%$ when temperature is employed as the thermodynamic
variable of ensemble extension \cite{Rathore05}).

Figure \ref{ar-ev} b) shows the evolution of the pressure versus
density plot with the number of performed trials during the
initializing procedure ($N=108$). For $\sim 10^{10}$ trials, the
sampling yields a curve at $\beta P > \beta P_{tr}$ which may
correspond to the random close packing (RCP) metastable branch.
There are also 5 replicas which reached the equilibrium state
(since they crystalized, they were pushed towards the highest
pressure region). Another one, producing the point laying on the
dotted line, may correspond to a partially crystalized structure.
Assuming this well defined curve corresponds to the RCP branch,
$\rho_{rcp}=0.680\pm0.005$ is obtained from $\beta P/\rho^2
\propto (\rho_{rcp}-\rho)^{-1}$ \cite{Rintoul96}. This is larger
than the reported value of $\rho_{rcp}=0.644\pm0.005$
\cite{Rintoul96}, suggesting that some degree of crystallization
is already taking place on the replicas at high pressure. This is
in fact confirmed by the $Q_6$ analysis (not shown). As the
initializing process advances, the degree of crystallization
augments and the high pressure curve shifts approaching the
equilibrium branch ($Q_6$ also enlarges). For $\sim 10^{12}$
trials, the curve practically yields the equilibrium branch. From
here on, only those replicas having a large degree of
crystallinity are able to access the high pressure region. At this
point, the initializing procedure ends and the sampling from
equilibrium process starts.


\begin{thebibliography}{10}%
\makeatletter
\providecommand \@ifxundefined [1]{%
 \ifx #1\undefined \expandafter \@firstoftwo
 \else \expandafter \@secondoftwo
\fi
}%
\providecommand \@ifnum [1]{%
 \ifnum #1\expandafter \@firstoftwo
 \else \expandafter \@secondoftwo
\fi
}%
\providecommand \enquote [1]{``#1''}%
\providecommand \bibnamefont  [1]{#1}%
\providecommand \bibfnamefont [1]{#1}%
\providecommand \citenamefont [1]{#1}%
\providecommand\href[0]{\@sanitize\@href}%
\providecommand\@href[1]{\endgroup\@@startlink{#1}\endgroup\@@href}%
\providecommand\@@href[1]{#1\@@endlink}%
\providecommand \@sanitize [0]{\begingroup\catcode`\&12\catcode`\#12\relax}%
\@ifxundefined \pdfoutput {\@firstoftwo}{%
 \@ifnum{\z@=\pdfoutput}{\@firstoftwo}{\@secondoftwo}%
}{%
 \providecommand\@@startlink[1]{\leavevmode}%
 \providecommand\@@endlink[0]{}%
}{%
 \providecommand\@@startlink[1]{%
  \leavevmode
  \pdfstartlink
   attr{/Border[0 0 1 ]/H/I/C[0 1 1]}%
   user{/Subtype/Link/A<</Type/Action/S/URI/URI(#1)>>}%
  \relax
 }%
 \providecommand\@@endlink[0]{\pdfendlink}%
}%
\providecommand \url  [0]{\begingroup\@sanitize \@url }%
\providecommand \@url [1]{\endgroup\@href {#1}{\urlprefix}}%
\providecommand \urlprefix [0]{URL }%
\providecommand \Eprint[0]{\href }%
\@ifxundefined \urlstyle {%
  \providecommand \doi [1]{doi:\discretionary{}{}{}#1}%
}{%
  \providecommand \doi [0]{doi:\discretionary{}{}{}\begingroup
  \urlstyle{rm}\Url }%
}%
\providecommand \doibase [0]{http://dx.doi.org/}%
\providecommand \Doi[1]{\href{\doibase#1}}%
\providecommand \bibAnnote [3]{%
  \BibitemShut{#1}%
  \begin{quotation}\noindent
    \textsc{Key:}\ #2\\\textsc{Annotation:}\ #3%
  \end{quotation}%
}%
\providecommand \bibAnnoteFile [2]{%
  \IfFileExists{#2}{\bibAnnote {#1} {#2} {\input{#2}}}{}%
}%
\providecommand \typeout [0]{\immediate \write \m@ne }%
\providecommand \selectlanguage [0]{\@gobble}%
\providecommand \bibinfo [0]{\@secondoftwo}%
\providecommand \bibfield [0]{\@secondoftwo}%
\providecommand \translation [1]{[#1]}%
\providecommand \BibitemOpen[0]{}%
\providecommand \bibitemStop [0]{}%
\providecommand \bibitemNoStop [0]{.\EOS\space}%
\providecommand \EOS [0]{\spacefactor3000\relax}%
\providecommand \BibitemShut [1]{\csname bibitem#1\endcsname}%
\bibitem{Lyubartsev91}%
  \BibitemOpen
  \bibfield{author}{%
  \bibinfo {author} {\bibfnamefont{A.~P.}\ \bibnamefont{Lyubartsev}}, \bibinfo
  {author} {\bibfnamefont{A.~A.}\ \bibnamefont{Martsinovski}}, \bibinfo
  {author} {\bibfnamefont{S.~V.}\ \bibnamefont{Shevkunov}},\ and\ \bibinfo
  {author} {\bibfnamefont{P.~N.}\ \bibnamefont{Vorontsov-Velyaminov}},\ }%
  \bibfield{journal}{%
  \bibinfo {journal} {J. Chem. Phys.}\ }%
  \textbf{\bibinfo {volume} {96}},\ \bibinfo {pages} {1776} (\bibinfo {year}
  {1991})%
  \bibAnnoteFile{NoStop}{Lyubartsev91}%
\bibitem{Marinari92}%
  \BibitemOpen
  \bibfield{author}{%
  \bibinfo {author} {\bibfnamefont{E.}~\bibnamefont{Marinari}}\ and\ \bibinfo
  {author} {\bibfnamefont{G.}~\bibnamefont{Parisi}},\ }%
  \bibfield{journal}{%
  \bibinfo {journal} {Europhys. Lett.}\ }%
  \textbf{\bibinfo {volume} {19}},\ \bibinfo {pages} {451} (\bibinfo {year}
  {1992})%
  \bibAnnoteFile{NoStop}{Marinari92}%
\bibitem{Yan99}%
  \BibitemOpen
  \bibfield{author}{%
  \bibinfo {author} {\bibfnamefont{Q.~L.}\ \bibnamefont{Yan}}\ and\ \bibinfo
  {author} {\bibfnamefont{J.~J.}\ \bibnamefont{de~Pablo}},\ }%
  \bibfield{journal}{%
  \bibinfo {journal} {J. Chem. Phys.}\ }%
  \textbf{\bibinfo {volume} {111}},\ \bibinfo {pages} {9509} (\bibinfo {year}
  {1992})%
  \bibAnnoteFile{NoStop}{Yan99}%
\bibitem{Bittner08}%
  \BibitemOpen
  \bibfield{author}{%
  \bibinfo {author} {\bibfnamefont{E.}~\bibnamefont{Bittner}}, \bibinfo
  {author} {\bibfnamefont{A.}~\bibnamefont{Nu{\ss}baumer}},\ and\ \bibinfo
  {author} {\bibfnamefont{W.}~\bibnamefont{Janke}},\ }%
  \bibfield{journal}{%
  \bibinfo {journal} {Phys. Rev. Lett.}\ }%
  \textbf{\bibinfo {volume} {101}},\ \bibinfo {pages} {130603} (\bibinfo {year}
  {2008})%
  \bibAnnoteFile{NoStop}{Bittner08}%
\bibitem{Frenkel}%
  \BibitemOpen
  \bibfield{author}{%
  \bibinfo {author} {\bibfnamefont{D.}~\bibnamefont{Frenkel}}\ and\ \bibinfo
  {author} {\bibfnamefont{B.}~\bibnamefont{Smit}},\ }%
  \emph{\bibinfo {title} {Understanding molecular simulation}}\ (\bibinfo
  {publisher} {Academic},\ \bibinfo {address} {New York},\ \bibinfo {year}
  {1996})%
  \bibAnnoteFile{NoStop}{Frenkel}%
\bibitem{Falcioni99}%
  \BibitemOpen
  \bibfield{author}{%
  \bibinfo {author} {\bibfnamefont{M.}~\bibnamefont{Falcioni}}\ and\ \bibinfo
  {author} {\bibfnamefont{M.~W.}\ \bibnamefont{Deem}},\ }%
  \bibfield{journal}{%
  \bibinfo {journal} {J. Chem. Phys.}\ }%
  \textbf{\bibinfo {volume} {110}},\ \bibinfo {pages} {1754} (\bibinfo {year}
  {1999})%
  \bibAnnoteFile{NoStop}{Falcioni99}%
\bibitem{Hernandez08}%
  \BibitemOpen
  \bibfield{author}{%
  \bibinfo {author} {\bibfnamefont{J.}~\bibnamefont{Hern\'{a}ndez-Rojas}}\ and\
  \bibinfo {author} {\bibfnamefont{J.~M.~G.}\ \bibnamefont{Llorente}},\ }%
  \bibfield{journal}{%
  \bibinfo {journal} {Phys. Rev. Lett.}\ }%
  \textbf{\bibinfo {volume} {100}},\ \bibinfo {pages} {258104} (\bibinfo {year}
  {2008})%
  \bibAnnoteFile{NoStop}{Hernandez08}%
\bibitem{Fiore08}%
  \BibitemOpen
  \bibfield{author}{%
  \bibinfo {author} {\bibfnamefont{C.~E.}\ \bibnamefont{Fiore}},\ }%
  \bibfield{journal}{%
  \bibinfo {journal} {Phys. Rev. E.}\ }%
  \textbf{\bibinfo {volume} {78}},\ \bibinfo {pages} {041109} (\bibinfo {year}
  {2008})%
  \bibAnnoteFile{NoStop}{Fiore08}%
\bibitem{Imperio06}%
  \BibitemOpen
  \bibfield{author}{%
  \bibinfo {author} {\bibfnamefont{A.}~\bibnamefont{Imperio}}\ and\ \bibinfo
  {author} {\bibfnamefont{L.}~\bibnamefont{Reatto}},\ }%
  \bibfield{journal}{%
  \bibinfo {journal} {J. Chem. Phys.}\ }%
  \textbf{\bibinfo {volume} {124}},\ \bibinfo {pages} {164712} (\bibinfo {year}
  {2006})%
  \bibAnnoteFile{NoStop}{Imperio06}%
\bibitem{Arnold08}%
  \BibitemOpen
  \bibfield{author}{%
  \bibinfo {author} {\bibfnamefont{A.}~\bibnamefont{Arnold}}\ and\ \bibinfo
  {author} {\bibfnamefont{C.}~\bibnamefont{Holm}},\ }%
  \bibfield{journal}{%
  \bibinfo {journal} {Eur. Phys. J. E}\ }%
  \textbf{\bibinfo {volume} {27}},\ \bibinfo {pages} {21} (\bibinfo {year}
  {2008})%
  \bibAnnoteFile{NoStop}{Arnold08}%
\bibitem{Fortini06}%
  \BibitemOpen
  \bibfield{author}{%
  \bibinfo {author} {\bibfnamefont{A.}~\bibnamefont{Fortini}}\ and\ \bibinfo
  {author} {\bibfnamefont{M.}~\bibnamefont{Dijkstra}},\ }%
  \bibfield{journal}{%
  \bibinfo {journal} {J. Phys.: Condens. Matter}\ }%
  \textbf{\bibinfo {volume} {18}},\ \bibinfo {pages} {L371} (\bibinfo {year}
  {2006})%
  \bibAnnoteFile{NoStop}{Fortini06}%
\bibitem{Odriozola08}%
  \BibitemOpen
  \bibfield{author}{%
  \bibinfo {author} {\bibfnamefont{G.}~\bibnamefont{Odriozola}}, \bibinfo
  {author} {\bibfnamefont{F.}~\bibnamefont{Jim\'{e}nez-\'{A}ngeles}},\ and\
  \bibinfo {author} {\bibfnamefont{M.}~\bibnamefont{Lozada-Cassou}},\ }%
  \bibfield{journal}{%
  \bibinfo {journal} {J. Chem. Phys.}\ }%
  \textbf{\bibinfo {volume} {129}},\ \bibinfo {pages} {111101} (\bibinfo {year}
  {2008})%
  \bibAnnoteFile{NoStop}{Odriozola08}%
\bibitem{Jimenez08}%
  \BibitemOpen
  \bibfield{author}{%
  \bibinfo {author} {\bibfnamefont{F.}~\bibnamefont{Jim\'{e}nez-\'{A}ngeles}},
  \bibinfo {author} {\bibfnamefont{Y.}~\bibnamefont{Duda}}, \bibinfo {author}
  {\bibfnamefont{G.}~\bibnamefont{Odriozola}},\ and\ \bibinfo {author}
  {\bibfnamefont{M.}~\bibnamefont{Lozada-Cassou}},\ }%
  \bibfield{journal}{%
  \bibinfo {journal} {J. Chem. Phys.}\ }%
  \textbf{\bibinfo {volume} {129}},\ \bibinfo {pages} {111101} (\bibinfo {year}
  {2008})%
  \bibAnnoteFile{NoStop}{Jimenez08}%
\bibitem{Wilding00}%
  \BibitemOpen
  \bibfield{author}{%
  \bibinfo {author} {\bibfnamefont{N.~B.}\ \bibnamefont{Wilding}}\ and\
  \bibinfo {author} {\bibfnamefont{A.~D.}\ \bibnamefont{Bruce}},\ }%
  \bibfield{journal}{%
  \bibinfo {journal} {Phys. Rev. Lett.}\ }%
  \textbf{\bibinfo {volume} {79}},\ \bibinfo {pages} {3002} (\bibinfo {year}
  {1997})%
  \bibAnnoteFile{NoStop}{Wilding00}%
\bibitem{Noya08}%
  \BibitemOpen
  \bibfield{author}{%
  \bibinfo {author} {\bibfnamefont{E.~G.}\ \bibnamefont{Noya}}, \bibinfo
  {author} {\bibfnamefont{C.}~\bibnamefont{Vega}},\ and\ \bibinfo {author}
  {\bibfnamefont{E.}~\bibnamefont{de~Miguel}},\ }%
  \bibfield{journal}{%
  \bibinfo {journal} {J. Chem. Phys.}\ }%
  \textbf{\bibinfo {volume} {128}},\ \bibinfo {pages} {154507} (\bibinfo {year}
  {2008})%
  \bibAnnoteFile{NoStop}{Noya08}%
\bibitem{Rosenbluth54}%
  \BibitemOpen
  \bibfield{author}{%
  \bibinfo {author} {\bibfnamefont{M.~N.}\ \bibnamefont{Rosenbluth}}\ and\
  \bibinfo {author} {\bibfnamefont{A.~W.}\ \bibnamefont{Rosenbluth}},\ }%
  \bibfield{journal}{%
  \bibinfo {journal} {J. Chem. Phys.}\ }%
  \textbf{\bibinfo {volume} {22}},\ \bibinfo {pages} {881} (\bibinfo {year}
  {1954})%
  \bibAnnoteFile{NoStop}{Rosenbluth54}%
\bibitem{Wood57}%
  \BibitemOpen
  \bibfield{author}{%
  \bibinfo {author} {\bibfnamefont{W.~W.}\ \bibnamefont{Wood}}\ and\ \bibinfo
  {author} {\bibfnamefont{J.~D.}\ \bibnamefont{Jacobson}},\ }%
  \bibfield{journal}{%
  \bibinfo {journal} {J. Chem. Phys.}\ }%
  \textbf{\bibinfo {volume} {27}},\ \bibinfo {pages} {1207} (\bibinfo {year}
  {1957})%
  \bibAnnoteFile{NoStop}{Wood57}%
\bibitem{Alder57}%
  \BibitemOpen
  \bibfield{author}{%
  \bibinfo {author} {\bibfnamefont{B.~J.}\ \bibnamefont{Alder}}\ and\ \bibinfo
  {author} {\bibfnamefont{T.~E.}\ \bibnamefont{Wainwright}},\ }%
  \bibfield{journal}{%
  \bibinfo {journal} {J. Chem. Phys.}\ }%
  \textbf{\bibinfo {volume} {27}},\ \bibinfo {pages} {1208} (\bibinfo {year}
  {1957})%
  \bibAnnoteFile{NoStop}{Alder57}%
\bibitem{Rintoul96}%
  \BibitemOpen
  \bibfield{author}{%
  \bibinfo {author} {\bibfnamefont{M.~D.}\ \bibnamefont{Rintoul}}\ and\
  \bibinfo {author} {\bibfnamefont{S.}~\bibnamefont{Torquato}},\ }%
  \bibfield{journal}{%
  \bibinfo {journal} {J. Chem. Phys.}\ }%
  \textbf{\bibinfo {volume} {105}},\ \bibinfo {pages} {9258} (\bibinfo {year}
  {1996})%
  \bibAnnoteFile{NoStop}{Rintoul96}%
\bibitem{Okabe01}%
  \BibitemOpen
  \bibfield{author}{%
  \bibinfo {author} {\bibfnamefont{T.}~\bibnamefont{Okabe}}, \bibinfo {author}
  {\bibfnamefont{M.}~\bibnamefont{Kawata}}, \bibinfo {author}
  {\bibfnamefont{Y.}~\bibnamefont{Okamoto}},\ and\ \bibinfo {author}
  {\bibfnamefont{M.}~\bibnamefont{Masuhiro}},\ }%
  \bibfield{journal}{%
  \bibinfo {journal} {Colloid Polym. Sci.}\ }%
  \textbf{\bibinfo {volume} {335}},\ \bibinfo {pages} {435} (\bibinfo {year}
  {2001})%
  \bibAnnoteFile{NoStop}{Okabe01}%
\bibitem{Rathore05}%
  \BibitemOpen
  \bibfield{author}{%
  \bibinfo {author} {\bibfnamefont{N.}~\bibnamefont{Rathore}}, \bibinfo
  {author} {\bibfnamefont{M.}~\bibnamefont{Chopra}},\ and\ \bibinfo {author}
  {\bibfnamefont{J.~J.}\ \bibnamefont{de~Pablo}},\ }%
  \bibfield{journal}{%
  \bibinfo {journal} {J. Chem. Phys.}\ }%
  \textbf{\bibinfo {volume} {122}},\ \bibinfo {pages} {024111} (\bibinfo {year}
  {2005})%
  \bibAnnoteFile{NoStop}{Rathore05}%
\bibitem{Steinhardt83}%
  \BibitemOpen
  \bibfield{author}{%
  \bibinfo {author} {\bibfnamefont{P.~J.}\ \bibnamefont{Seinhardt}}, \bibinfo
  {author} {\bibfnamefont{D.~R.}\ \bibnamefont{Nelson}},\ and\ \bibinfo
  {author} {\bibfnamefont{M.}~\bibnamefont{Ronchetti}},\ }%
  \bibfield{journal}{%
  \bibinfo {journal} {Phys. Rev. B.}\ }%
  \textbf{\bibinfo {volume} {28}},\ \bibinfo {pages} {784} (\bibinfo {year}
  {1996})%
  \bibAnnoteFile{NoStop}{Steinhardt83}%
\bibitem{Speedy97}%
  \BibitemOpen
  \bibfield{author}{%
  \bibinfo {author} {\bibfnamefont{R.~J.}\ \bibnamefont{Speedy}},\ }%
  \bibfield{journal}{%
  \bibinfo {journal} {J. Phys.: Condens. Matter}\ }%
  \textbf{\bibinfo {volume} {9}},\ \bibinfo {pages} {8591} (\bibinfo {year}
  {1997})%
  \bibAnnoteFile{NoStop}{Speedy97}%
\bibitem{Speedy98}%
  \BibitemOpen
  \bibfield{author}{%
  \bibinfo {author} {\bibfnamefont{R.~J.}\ \bibnamefont{Speedy}},\ }%
  \bibfield{journal}{%
  \bibinfo {journal} {J. Phys.: Condens. Matter}\ }%
  \textbf{\bibinfo {volume} {10}},\ \bibinfo {pages} {4387} (\bibinfo {year}
  {1998})%
  \bibAnnoteFile{NoStop}{Speedy98}%
\end{thebibliography}
\end{document}